\documentstyle[epsf,12pt]{article}
\catcode`@=11
\long\def\@caption#1[#2]#3{\par\addcontentsline{\csname
  ext@#1\endcsname}{#1}{\protect\numberline{\csname
  the#1\endcsname}{\ignorespaces #2}}\begingroup
    \small
    \@parboxrestore
    \@makecaption{\csname fnum@#1\endcsname}{\ignorespaces #3}\par
  \endgroup}
\catcode`@=12
\newcommand{\newc}{\newcommand}
\newc{\gsim}{\lower.7ex\hbox{$\;\stackrel{\textstyle>}{\sim}\;$}}
\newc{\lsim}{\lower.7ex\hbox{$\;\stackrel{\textstyle<}{\sim}\;$}}
\newc{\gev}{\,{\rm GeV}}
\newc{\mev}{\,{\rm MeV}}
\newc{\mmev}{\,{\rm meV}}
\newc{\ev}{\,{\rm eV}}
\newc{\kev}{\,{\rm keV}}
\newc{\tev}{\,{\rm TeV}}
\newc{\mz}{m_Z}
\newc{\mpl}{M_{Pl}}
\newc{\chifc}{\chi_{{}_{\!F\!C}}}
\newc\order{{\cal O}}
\newc\CO{\order}
\newc\CL{{\cal L}}
\newc\CY{{\cal Y}}
\newc\CH{{\cal H}}
\newc\CM{{\cal M}}
\newc\CF{{\cal F}}
\newc\CD{{\cal D}}
\newc\CN{{\cal N}}
\newc{\eps}{\epsilon}
\newc{\re}{\mbox{Re}\,}
\newc{\im}{\mbox{Im}\,}
\newc{\invpb}{\,\mbox{pb}^{-1}}
\newc{\invfb}{\,\mbox{fb}^{-1}}
\newc{\yddiag}{{\bf D}}
\newc{\yddiagd}{{\bf D^\dagger}}
\newc{\yudiag}{{\bf U}}
\newc{\yudiagd}{{\bf U^\dagger}}
\newc{\yd}{{\bf Y_D}}
\newc{\ydd}{{\bf Y_D^\dagger}}
\newc{\yu}{{\bf Y_U}}
\newc{\yud}{{\bf Y_U^\dagger}}
\newc{\ckm}{{\bf V}}
\newc{\ckmd}{{\bf V^\dagger}}
\newc{\ckmz}{{\bf V^0}}
\newc{\ckmzd}{{\bf V^{0\dagger}}}
\newc{\X}{{\bf X}}
\newc{\bbbar}{B^0-\bar B^0}

\newc{\sgn}{\mbox{sgn}\,}
\newc{\m}{{\bf m}}
\newc{\msusy}{M_{\rm SUSY}}
\newc{\munif}{M_{\rm unif}}
\relax
\def\beq{\begin{equation}}
\def\eeq{\end{equation}}
\def\bea{\begin{eqnarray}}
\def\eea{\end{eqnarray}}
\newc{\ie}{{\it i.e.}}          \newc{\etal}{{\it et al.}}
\newc{\eg}{{\it e.g.}}          \newc{\etc}{{\it etc.}}
\newc{\cf}{{\it c.f.}}
\def\Dsl{\,\raise.15ex\hbox{/}\mkern-13.5mu D} 
\def\delsl{\raise.15ex\hbox{/}\kern-.57em\partial}
\def\Ksl{\hbox{/\kern-.6000em\rm K}}
\def\Asl{\hbox{/\kern-.6500em \rm A}}
\def\Qsl{\hbox{/\kern-.6000em\rm Q}}
\def\gradsl{\hbox{/\kern-.6500em$\nabla$}}
\def\bar#1{\overline{#1}}

\title{\textbf{ Quintessence and Varying $\alpha$ from Shape Moduli}}
\author{Mark Byrne and Christopher Kolda\\ \\
{\it Department of Physics, University of Notre Dame}\\
{\it Notre Dame, IN 46556, USA}}
\date{}

\begin{document}

\maketitle

\begin{abstract}
In extra-dimensional models which are compactified on an $n$-torus
($n\geq 2$) there exist moduli associated with the torus volume (which
sets the fundamental Planck scale), the ratios of the torus radii, and
the angle(s) of periodicity. We consider a
model with gravity in the bulk of $n=2$ 
large extra dimensions with a fixed volume, 
taking all Standard Model fields to be 
confined to a ``thick" and supersymmetric 3-brane.
The Casimir energy of fields in the bulk of the 2-torus 
accounts for the present dark
energy density while the shape moduli begin rolling at late times ($z
\sim 1$) and induce 
a shift in the Kaluza-Klein masses of the Standard Model fields. 
The low energy value of the fine-structure constant is
sensitive at loop level to this shift. For reasonable cosmological
initial conditions on the shape moduli we obtain a redshift dependence of
the fine-structure constant similar to that reported by Webb \etal,  which
is roughly compatible with Oklo and meteorite bounds.  
Constraints from coincident variation in the QCD scale are also briefly discussed.

\end{abstract}

\newpage

\setcounter{footnote}{0}
\setcounter{page}{1}
\setcounter{section}{0}
\setcounter{subsection}{0}
\setcounter{subsubsection}{0}


\section{Introduction}

In recent years the quality and quantity of cosmological and
astrophysical data has improved to the point that
a precise determination of the fundamental cosmological parameters
is now possible.  With independent estimates of the cosmological
parameters from a variety of sources (WMAP, type Ia supernovae,
large-scale structure, etc.), a compelling picture has emerged
suggesting that we live in a flat universe: dark
energy comprises roughly 70\% of the total energy density
of the universe, while the remainder is primarily cold, non-baryonic dark
matter~\cite{papers2}. With increasing astrophysical data at high
redshifts it is also possible to test and constrain
temporal deviations of the fundamental constants.
There has been a reported variation of the fine-structure
constant, $\alpha$, in the range of redshifts $z \approx 0.5-3$. The 
observed deviation, if averaged over the
same timescale, is $4.7\sigma$ from zero: $\delta
\alpha/\alpha = (0.543\pm0.116) \times 10^{-5}$~\cite{webb1}.
We will attempt to explain this deviation, 
although there are alternative interpretations of the observations
which do not require a
time-varying $\alpha$~\cite{papers1}. 

A variation in $\alpha$ does seem natural from the perspective of string
theory, since all coupling constants are in principle set by vacuum
expectation values (vevs) of scalar fields. However, one expects
that these fields were stabilized at very early times, $t \sim
1/\mpl$ or at least prior to Big-Bang nucleosynthesis, and
that their masses are of order the Planck scale.
A variation in a gauge coupling at low redshift contradicts
this expectation, suggesting that there is at least one moduli which has 
only recently (or not yet) reached its global minimum. 

Heuristically, this sounds very similar to quintessence models which
seek to explain the cosmological acceleration by a slowly rolling
scalar field.
Thus it might seem natural to try to connect models of varying $\alpha$ with
models of quintessence. There is even an added bonus when trying to
connect these two ideas:
the epoch from which the data on varying $\alpha$ is taken
is also the epoch during which the quintessence would have begun
to dominate the energy density of the universe. Attempts have been made to
create a unified picture of quintessence and varying $\alpha$ that
make use of this coincidence~\cite{papers3, goldberg}.

However there is a more serious problem. In order to maintain their
extraordinarily flat potentials, and in order to avoid large violations
of the equivalence principle~\cite{carroll,kolda}, 
quintessence fields must have no (or only
infinitesimal) couplings to ordinary matter. But the field that
generates time-varying $\alpha$ must by necessity have non-negligible
couplings to either photons or to electrically-charged matter.

In this paper we will use supersymmetry and extra dimensions
together to solve the problems that plague models of quintessence and
varying $\alpha$, following the earlier work of Peloso and
Poppitz~\cite{poppitz} and Pietroni~\cite{pietroni}, who considered
quintessence alone. 
Both papers made use of a simple result: in a
scenario with two large, flat extra dimensions and a TeV gravity scale, 
the compactification volume is 
$M_{Pl} ^2 M_*^{-4}= V_{(2)}$ which has a mass scale $V_{(2)}^{-1/2}\sim
{\cal O}(\mmev)$.  If we identify the quintessence field with a
shape modulus of the compactification manifold, then the
finite contributions to the
self-energy of bulk fields are ${\cal O}(\mmev^4)$, which is the same
order of magnitude as the current dark energy density. 

We will flesh out our model in the next few sections, but 
we can summarize it as follows: the quintessence field is
identified with a shape modulus of a 2-torus compactification manifold
whose volume is kept constant. The Casimir energy
of fields in the bulk, whose masses are naturally ${\cal O}(V^{-1/2})$, 
provide the dark energy density. Assuming
natural initial conditions on the shape moduli, the fields are frozen
for most of the history of the universe and only very recently ($z \sim 1$)
began rolling toward the global minimum of their potential. (This is
essentially the model in Ref.~\cite{poppitz}.) 
To relate this to a variation in $\alpha$,  
we suppose that the 3-brane on which the SM lives has a finite thickness
out into the bulk. The SM ``fat brane'' then has a Kaluza-Klein (KK) 
tower of states whose masses are sensitive to the global topology of
the bulk, and in particular, to the angle of periodicity of the 2-torus.

More explicitly, 
we make the usual assumption that the physics that sets the size of
$\alpha$ sits at the gravitational cut-off, $M_*$, as one would expect
in string theory. The low scale value of $\alpha$ is then derived
through use of the renormalization group, which is sensitive at one
loop to the masses of the KK excitations which lie between the weak
and gravitational cut-off scales (see, \eg, Ref.~\cite{chacko}).
Therefore $\alpha$ is sensitive to the topology of the compactification
manifold. By itself such a model nicely explains
the time evolution of $\alpha$ and is roughly consistent with all known
constraints. But in order to simultaneously explain the quintessence,
we must assume that the spectrum of excited KK states 
on the fat brane is supersymmetric; otherwise the Casimir
energy associated with the fat brane is ${\cal O}(M_W^4)$ and 
dominates the cosmological constant. In such a case we expect that
boundary conditions are responsible for breaking SUSY so that
the zero modes are identified with the SM sector and the nonzero modes 
form supersymmetric multiplets with degenerate masses.

For reasonable initial conditions on the moduli we find a variation of
$\alpha$ similar to that reported in the literature~\cite
{webb1}. The predicted deviation is roughly compatible with  the
constraints from Oklo ($z = 0.14$) and meteorite ($z = 0.45$)
bounds. Section~2 discusses the model in detail, the KK masses of the fields,
and briefly summarizes experimental constraints for asymmetric tori
with a shift angle. In section~3 we evaluate the change in the low
energy value of $\alpha$ from the 1-loop renormalization group
equations as a function of the
fractional shift in the masses and briefly comment on possible correlations 
with the time variation of $\Lambda_{QCD}$. In section 4 
we solve for the motion
of the moduli in an FRW universe and evaluate the time evolution of
$\alpha$.

\section{General Setup and Toroidal Compactification}

We will work in a six dimensional spacetime with coordinates $x^M =
\{x^{\mu}, y_1,y_2\}$ where $\mu$ runs over the usual four dimensional
indices and $y_a$ are the extra spatial coordinate labels.  The two
large extra dimensions are compactified on a torus with a shift angle
$\theta$ between the axes of periodicity (see Fig.~\ref{fig1}). 
Let $R_1$ and $R_2$ denote
the radii of the torus and such that $R_2 \geq R_1$.
We parameterize the torus following Ref.~\cite{dienes}:     
\begin{figure}
\centering
\hspace*{0in}
\epsfxsize=3.00in
\epsffile{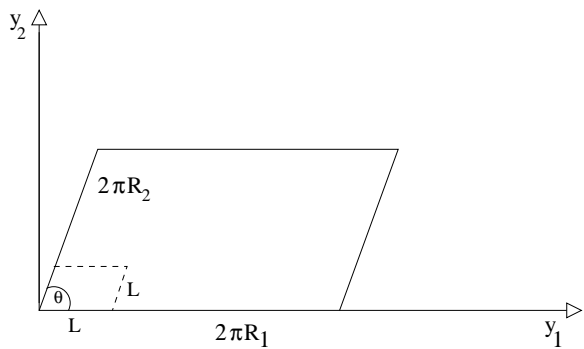}
\caption{Two large extra dimensions compactified on a torus with the
Standard Model fields confined to a subspace of area $L^2 << 4 \pi ^2
R_1 R_2 \sin{\theta}$.
} 
\label{fig1}
\end{figure}
\begin{eqnarray}
A &=& 4 \pi ^2 R_1 R_2 \sin{\theta}, \\  
\tau &=& \frac{R_2}{R_1} e^{i \theta} \equiv \tau_1 + i\tau_2 
\end{eqnarray}
where $A$ is the area (or volume) modulus and $\tau$ is a complex
shape modulus.
The background metric for the spacetime is of the form:
\begin{equation}
ds^2 = g^{\mu \nu} dx_{\mu} dx_{\nu}  + h^{a b} dy_a dy_b,
\end{equation}
where $y_1 \in (0, 2\pi R_1)$ and $y_2 \in (0, 2\pi R_2
\sin{\theta})$.  The metric on the torus can be written as~\cite{poppitz}: 
\begin{equation}
h^{ab} = \frac{1}{\tau_2}
\left( \begin{array}{cc} 1 & \tau_1 \\
\tau_1 & |\tau|^{2} \end{array} \right).
\end{equation}

The SM fields are, by assumption, confined to a 3-brane with
some finite extent into the bulk. We obtain the four dimensional effective
action by integrating over the two extra coordinates. Due
to our ignorance of the confinement mechanism we can simply
restrict the SM sector to a subspace of the torus by a product of step
function terms in the action. However, any smooth functions which
generate the confinement should work equally well in this
analysis. Thus we can write an action: 
\begin{equation}     
S = M_* ^4 \int d^4x\, d^2y\, \sqrt{-G}\,\left\{R(G) +
\cdots ~+\left(\bar{\Psi}\Gamma^M D_M\Psi + \cdots\right)
\Theta(L -y_1)\Theta(L-y_2)\right\}. 
\end{equation}
The first ellipses represents fields in addition to gravity
which propagate in the bulk while the second ellipses 
represents the SM sector confined to the fat brane $0\leq y_{1,2}\leq
L$. (We will assume for simplicity that the thick brane is symmetric
in the 2 extra dimensions.) 
The fundamental Planck scale is defined by 
\beq
\mpl^2=4
\pi ^2 R_1 R_2 \sin{\theta} M_*^4
\eeq
which we hold constant. 
Due to non-observation of direct KK
production or virtual exchange of KK Standard Model modes, the
confinement length is constrained to be $L\lsim\tev^{-1}$.
Note that the total area ($A$) of the
torus is $\sim \mmev^{-2}$ in order to give  an
$M_*\sim\tev$ gravitational scale.

The original eigenfunctions for fields in the bulk are determined by
the requirement that the fields are single-valued under the coordinate
identifications of the torus: $(y_1,y_2) \rightarrow (y_1 +2 \pi R_2
\cos{\theta}, ~y_2 +2 \pi R_2 \sin{\theta})$, $(y_1,y_2) \to
(y_1+ 2\pi R_1, y_2)$.  This implies for any field $\psi$ in the bulk:  
\begin{equation}
\psi(y_1,y_2) \propto \exp{~i\left( \frac{ n_1 y_1} {R_1} +
\frac{\hat{n}_2 y_2} {R_2 \sin{\theta}}\right) }
\end{equation}
where $n_1$ and $n_2$ are integers~\cite{dienes} and
\beq
\hat{n}_2  = n_2 \left\{1-\left(\frac{n_1}{n_2}\right) 
\left(\frac{R_2} {R_1}\right)
\cos{\theta} \right\}.
\eeq

For the Standard Model fields which are confined to the thick brane
there is one complication. 
In order to obtain the correct chiral zero mode
structure for the fermions, we can apply Dirichlet or Neumann
conditions on the fields at the boundaries of the subspace rather than
at the boundaries of the bulk, 
or equivalently we must impose an additional reflection symmetry $(y_1,y_2)
\rightarrow (-y_1,-y_2)$ assigning fields with even or odd parity
under this $Z_2$~\cite{ued}. 
Then the induced metric on the thick 3-brane~(\cite{sundrum}) implies that
fields in the subspace have similar eigenfunctions to the bulk
fields except that these are either even or odd. For example, a gauge
field has the expansion:
\begin{eqnarray} 
A_{\mu} (x^{\nu},y_i) &=& \sqrt{ \frac{2} {L^2} } 
\left\{A_{\mu}^{(0,0)}(x^{\nu}) + \sqrt{2} \sum_{n_1, n_2}
 A_{\mu} ^{(n_1,n_2)}(x^{\nu}) \cos{\left(\frac{2\pi n_1 y_1}{L}
 + \frac{2\pi\tilde{n}_2 y_2}{L}\right)}\right\}.
\end{eqnarray}
There is an analogous KK expansion for fermions
whose non-zero modes are vector-like~\cite{ued}.  More importantly,
at tree level and excluding electroweak symmetry-breaking effects,
all the KK masses are degenerate since they are determined solely 
from the boundary conditions.

The masses of the Standard Model KK modes have a functional
dependence on $\theta$ from  $\tilde{n}_2 = n_2 
(1-(n_1/n_2)\cot\theta)$. For gauge fields propagating in a square
torus the usual mode expansion follows by replacing $L = 2 \pi R$
and $\theta = \pi/2$.  Note that the (dimensionless) zero mode gauge
couplings are $g_4 = g_6/L$. Thus as the shape moduli of the bulk
roll, we must require that the width, $L$, of the brane stays
constant; otherwise $\alpha$ would change too rapidly. We
assume that whatever dynamics sets the area of the subspace in which
the SM sector lives, the area itself 
does not depend on the bulk shape moduli. (It
could depend on the bulk volume moduli, which is kept fixed.)

There is at least one potentially disastrous consequence of a rolling
modulus which alters the KK mass spectrum of fields in the subspace. On
general grounds the contribution to the effective
cosmological constant from a
rolling modulus (in this case the angle) in the subspace will be of
order 
$L^{-4}\,(\delta M/M)$ where $M$ is the mass scale associated
with KK states and $\delta M$ is a typical shift in the masses from
the motion of the moduli. Since $L \sim \tev^{-1}$ the change in the
effective cosmological constant would be huge unless the typical
fractional shift in the masses were very small, $\delta M/M
\sim 10^{-15}$. As we will see, 
a shift in the masses of $O(10^{-3})$ is needed to accommodate
the varying-$\alpha$ data.  This contribution from dynamical KK modes to
the effective cosmological constant, however, is zero if the theory is
supersymmetric above the scale $1/L$ and we will assume this for the
remainder of the paper. By incorporating exact supersymmetry above the
inverse thickness scale, the only dynamical contribution to an
effective cosmological constant is from fields in the bulk of the
torus.     
 
\subsection{Graviton and Standard Model KK Masses}

First we consider the gravitational sector.  With the area of the
torus held constant the KK graviton masses can be expressed in 
terms of the ratio of the radii ($|\tau|$) and the angle $\theta$:
\begin{equation}
M_{n_1,n_2} ^2 = \frac{4 \pi^2}{A} \frac{|\tau|}{\sin{\theta}}\left(n_1
^2  +\frac{n_2 ^2}{|\tau|^2} -\frac{2 n_1 n_2}{|\tau|} \cos{\theta}\right)  
\label{eq:mass}
\end{equation}
where $n_1$ and $n_2$ are integers. 
For simplicity we will absorb the
constant factor $4\pi^2/A$ into Eq.~(\ref{eq:mass}).

There are two distinct
classes of constraints which place
limits on the masses of the KK gravitons~\cite{nima}. The first come from
processes in which real gravitons are produced; these limits are 
sensitive to the multiplicity of the KK
states below the relevant energy or temperature of the process being studied.
The second class arise from searches  
for deviations in the Newtonian potential at short 
distances. This
latter constraint is dominated by the mass gap of the lightest KK
graviton. We will calculate this mass gap now in order to check
the consistency of our model with experiment.

In order to calculate the KK graviton mass gap from
Eq.~(\ref{eq:mass}), we need only consider two cases: $\cos\theta=0$
or not. If $\cos\theta=0$, then the lightest KK mass is obtained for 
$n_1=0$ and $n_2=1$ (recall that $|\tau|\ge 1$).
Then the lightest KK state has mass $1/|\tau|=1/\tau_2$.
A KK graviton mass less than $O(\mmev)$ is
inconsistent with short-distance tests of Newton's laws (see
Ref.~\cite{pricelong} for reviews). However we will see in Section 4
that both extrema of the Casimir potential lie
near $\tau_2 \sim 1$. As long as the shape moduli are in the
neighborhood of their minima, deviations to the gravitational potential
will be consistent with experiment.

A mass gap which is inconsistent with experiment can also arise for
$\cos\theta\neq 0$ if there is a cancellation between the various
terms in Eq.~(\ref{eq:mass}). In order to examine this case more
carefully, 
take $|\tau|$ to be irrational (as it generally should
be) and write it as a rational part plus an irrational part of unit
modulus: $|\tau| =
|\tau|_r + |\tau|_{ir}$ with $|\tau|_{ir} \le 1$.  The lightest
possible states correspond to $ n_2 = |\tau|_r\, n_1$ where the masses
are 
\begin{equation}
M_n ^2 \sim n_1 ^2 \frac{|\tau|} { \sin{\theta} } \left\{ \left(
\frac{|\tau|_{ir}}{|\tau|} \right) ^2 + 2\left(1 -\cos{\theta}\right) 
\right\}. 
\end{equation}
The mass gap approaches zero when $|\tau|_{ir}\to 0$ {\it and}\/
$\cos\theta\to 1$. While such a model would be ruled out by searches
for deviations in Newton's force law, neither the limit
$|\tau|_{ir}\to 0$ nor $\cos\theta\to 1$ are preferred by the dynamics
of the model. Over the great majority of the parameter space in
$\theta$, or for typical $|\tau|_{ir}$, the mass gap is ${\cal
  O}(\mmev)$ and thus consistent with experiment. So
the bounds on the fundamental scale from deviation of the
Newtonian potential are typically the same as for the rectangular
symmetric torus, except in small regions of parameter space.

The other class of constraints come from direct production of KK
gravitons or by exchange of virtual gravitons in SM processes. The
strongest constraints come from stellar production in which the
gravitons are produced on-shell if their masses are below the typical
temperatures in the stellar interior.
In models with two extra dimensions compactified on a square torus,
constraints from supernova cooling imply $M_* \gsim 50 \tev$ ~\cite{cullen}.  
Given some maximum
temperature $T$, Eq.~(\ref{eq:mass}) under the constraint
$M^2_{n_1,n_2}\leq T^2$ defines an ellipse of
constraint whose area approximates the number of graviton states below
$T$; this area is independent of both the shift angle and the ratio of
radii.  Therefore, most collider,
astrophysical, and cosmological constraints on the fundamental scale
apply equally well to asymmetric tori as they do for rectangular tori, 
even though the two induce different mass spectra for the KK 
gravitons~\cite{byrne}.

For states which are charged under the SM symmetries, the KK mass
spectrum has a similar form as for the gravitons, though the relevant
mass scale is much larger:
\begin{equation}
{\cal{M}}_{n_1,n_2} ^2 = \frac{4 \pi^2}{L^2 \sin^2{\theta}} 
\left(n_1 ^2
+ n_2 ^2 \sin^2{\theta} -2 n_1 n_2 \sin{\theta} \cos{\theta}\right).
\label{kkmtheta}
\end{equation}
For $\theta = \pi /2 $ we obtain the spectrum usually discussed in the
literature.  The
mass scale $2\pi/L\sin\theta$ must be $\gsim {\cal O}(\tev)$
so that the KK states would have avoided detection;
a discussion of possible collider phenomenology is beyond the
scope of the paper and we refer the reader to Refs.~\cite{fatbrane,nandi}.
For future reference note that that
Eq.~(\ref{kkmtheta}) can be written in terms of the global moduli: 
\begin{equation}
\label{kkm}
{\cal{M}}_{n_1,n_2} ^2 = \frac{4 \pi^2}{L^2} \left
(\frac{|\tau|}{\tau_2}\right)^2   
\left[n_1 ^2  + n_2 ^2 \left( \frac{\tau_2}{|\tau|} \right)^2 -2 n_1
n_2 \frac{\tau_1 \tau_2} {|\tau| ^2} \right].
\end{equation}
In section 4 we will evaluate the change in $\tau_1$ and
$\tau_2$ from the relevant equations of motion.

\section{Rolling Moduli and Threshold Corrections}

It is apparent from the mass spectrum of Eq.~(\ref{kkmtheta}) 
that a
change in the shift angle $\theta$ will shift the 
masses of the Standard Model KK fields. In this section we will examine
the effect of changing the KK masses on the low energy value of the
gauge couplings. Since we fix the thickness of the brane, the gauge
couplings are only sensitive at loop level to changes in the KK masses.
In particular, the
fine structure constant is sensitive to changes in the KK masses of
electrically charged states.   

The quasar emission and absorption line data are sensitive to the
value of the fine-structure constant as measured at $q_0^2\sim\ev^2$.
However in theories of fundamental physics, the value of $\alpha$
is set by conditions in the ultraviolet, at some cut-off scale
$\Lambda$. In order to go from a
fundamental $\alpha(\Lambda)$ to a measured $\alpha(q_0)$, we use
the renormalization group equations (RGEs)
(see \cite{chacko} for a similar analysis):
\begin{equation}
\frac{1}{\alpha (q_0)} = \frac{1}{\alpha(\Lambda)} + \frac{1}{2 \pi}
\sum_{n=0} b_{n}\,{\log \frac{m_{n+1}}{m_n} }
\end{equation}
at one loop. 
In the above RGE, $m_n$ is the mass of the $n$-th KK excitation of the
charged SM fields, with $n=0$ representing the zero modes. 
(In this section, we will consider only one extra dimension,
generalizing to two in Section 5.)
Each KK
level only contributes to the RGEs at scales greater than its
mass. With the exception of the zero modes, each KK level contributes
equally to the $\beta$-functions, an amount $\tilde b$. Thus the
$\beta$-function coefficient between the $n$ and $(n+1)$ KK levels is
given by
\begin{equation}
b_{n} = b^{SM} + n \tilde{b}.
\label{beta}
\end{equation}
The infrared value of $\alpha$ is related to the charged KK masses by
\begin{equation}
\frac{1}{\alpha(q_0)} = \frac{1}{\alpha(\Lambda)} + \frac{b_{SM}}{2
\pi}{\log \frac{\Lambda}{q_0}} +  \frac{\tilde b}{2 \pi} \sum_{n=1} ^{N}
 n\, \log \frac{m_{n+1}}{m_n},
\end{equation}
where $m_{N+1}$ should be understood to be equal to $\Lambda$; that is,
the summation is terminated when the KK towers reach the ultraviolet
cut-off for the theory. The number of levels, $N$, in the KK tower 
will not be
a particularly large number, since the KK states will cause the gauge
couplings to run very quickly, reaching a Landau pole at scales below
$\Lambda$ unless $N$ is ${\cal O}(10)$ or less. A somewhat stronger
limit is obtained by demanding tree-level unitarity up to
$\Lambda$~\cite{chivukula}. 

We are interested in the fractional change in $\alpha$ caused by
varying the KK masses while the infrared ($q_0$) and ultraviolet
($\Lambda$) scales are held constant:
\begin{equation}
\frac{\delta \alpha}{\alpha} = \frac{\alpha}{2 \pi}\tilde b \sum_{n=1}^{N} 
\frac{\delta {\cal{M}}_n} {{\cal{M}}_n}. 
\label{alpha}
\end{equation}

In the simplest model, all of the SM fields will live on the thick
brane, and so there will be correlations between the shift in
$\alpha$ and corresponding shifts in the strong and weak coupling
constants. Following
Ref.~\cite{langacker} the variation in the QCD scale is related to
the variation in $\alpha$ as: 
\begin{equation}
\frac{\delta \Lambda_{QCD} }{\Lambda_{QCD}} = \frac{2 \pi}{9
\alpha} \,\frac{\tilde{b}_3}{\tilde{b}_1 +\tilde{b}_2}\,
\frac{\delta \alpha} {\alpha} \equiv \zeta \frac{\delta
\alpha} {\alpha}.
\end{equation}
The value of $\zeta$ is highly model-dependent. For example, the
strongly-interacting states could be confined to a ``thin'' brane or
orbifold fixed points so
that there are no colored KK modes for the SM fields and $\zeta=0$. 
Many other possibilities exist. Because the QCD
constraints require imposing another level of model dependence, we
will not consider them here.

\section{Cosmological Dynamics}

As we have seen in the previous sections, and as was first pointed out
by Dienes~\cite{dienes}, the masses of the bulk KK excitations are
functions of the torus shape moduli. Therefore an effective potential
is generated for the shape moduli, which can also be thought of as the
Casimir potential between the sides of the torus. This potential has been
explicitly calculated by Ponton and Poppitz~\cite{ponton}. Those
authors found that if the bulk contains gravity alone, the effective
potential for the shape moduli would be 
negative (anti-de Sitter space). Thus it is necessary to
put some fermions in the bulk to change the sign of the potential and
thereby generate a deSitter solution. Ref.~\cite{ponton} suggested
using neutrinos in the bulk, but an identification is not really
necessary. We will likewise assume that the spectrum in the
bulk contains sufficient fermions to yield $V\geq0$.

\begin{figure}
\centering
\hspace*{0in}
\epsfysize=2.5in
\epsffile{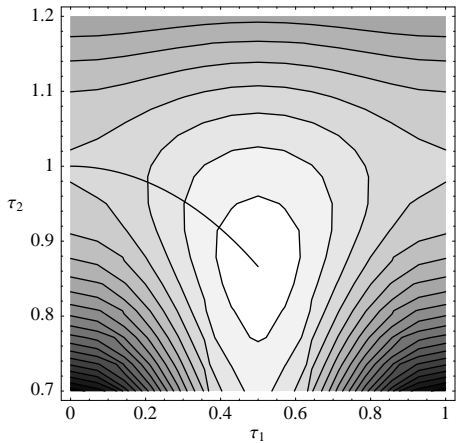}
\caption{
Contour plot of the Casimir potential 
in the $\tau_1$ -- $\tau_2$ plane. The saddle point where the moduli
begin rolling are at $(0,1)$ or $(1,1)$. The minimum is at
$(1/2,\sqrt{3}/2)$.}
\label{fig2}
\end{figure}

Fig.~\ref{fig2} shows a contour plot of the potential in
the region near the extrema and in the range $\tau_1 \in [0,1)$. The
potential has the modular invariance of the torus and is thus
periodic under $\tau_1\to\tau_1+1$. It has a general form:
\begin{equation}
V_C = \frac{1}{4 \pi ^2 R_1 R_2 \sin{\theta}}\, g(\tau_1, \tau_2).
\end{equation}
The volume prefactor sets the scale for quintessence, and $g$
is a function which is ${\cal O}(1)$ near the extrema of the potential. 
The two extrema (the saddle point and the minimum)
correspond to the modular invariants of the torus. In our model we
will assume that the shape modulus begins its evolution at early times
stuck near the saddle point where the potential has the
approximate form (there is little dependence on $\phi_2$):
\beq
 V \simeq V_0 \cos{\frac{\sqrt{2}\pi 
\phi_1}{\mpl}}
\eeq  
where $V_0$ is ${\cal O}(\mmev^4)$. We can then study its cosmological
evolution.

The action for the moduli in an FRW universe has a non-canonical
kinetic term: 
\begin{equation}
S = \frac{1}{2} M_{Pl} ^2 \int d^4x\, \left[a(t)\right]^3 
\left\{\frac{1}{2 \tau_2 ^2}
\left(\dot{\tau_1}^2 + \dot{\tau_2}^2\right) + V(\tau_1, \tau_2) \right\}.  
\end{equation}
The moduli fields can be defined in terms of the shape parameters:
\beq
\phi_1 \equiv \frac{\mpl}{\sqrt{2}}\,\tau_1, \quad\quad
\phi_2 \equiv \frac{\mpl}{\sqrt{2}}\,\log{\tau_2}. 
\eeq

The equation of motion for $\phi_1$ (coupled to $\phi_2$) is then
\begin{equation}
\ddot{\phi_1} + 3 H \dot{\phi_1} 
-\frac{2 \sqrt{2}}{\mpl}\, \dot{\phi_1} \dot{\phi_2}
=  \left(-\frac{\partial{V}}{\partial{\phi_1}}\right)\, \exp\frac{2\sqrt{2}
\,\phi_2}{\mpl}.  
\label{eq:motion}
\end{equation}
There is an analogous equation for $\phi_2$. Since we are
interested in the behavior of the field near the saddle point of the
potential, we take 
$\dot{\phi_2} \sim 0$ and the exponential in 
Eq.~(\ref{eq:motion}) to be about
one. This can be solved numerically by transforming the
coordinates $x \equiv \log{a} = -\log{(1+z)}$ (see \cite{goldberg},
for example) which gives an equation of motion: 
\begin{equation}
(V+\rho)\, \phi_1'' +  3\, (V+\rho)\,\left(1 - \phi_1'{}^2
/6\right)\phi_1' + \left(\rho'\phi_1'/2 + 3\,
\partial_\phi V\right)\,\left(1 - \phi_1'{}^2/6\right) = 0. 
\label{cosmo}
\end{equation}
In the above, $\rho$ is the contribution of matter and
radiation to the energy density
and a prime indicates a derivative with respect to $x$. 

We need to solve Eq.~(\ref{cosmo}) for the time evolution of $\tau_1$
given some reasonable ``initial'' conditions for the field at early
redshifts. For this particular form of quintessence the equation of
state is almost precisely that of a cosmological constant; that is,
$\phi_1$ barely rolls over the lifetime of the universe and
$w=p/\rho=-1$ to a high accuracy. Thus we simply require that 
$V_0 =0.7 \rho_0$ where
$\rho_0 = 1.05 \times 10^{-5} h^2 \gev\,\mbox{cm}^{-3}$ is the present
critical density. The contribution from matter
scales as $\rho_M = 0.3 \rho_0\,e^{-3x}$ where $x =
-\ln(1+z)$. In these variables, big bang nucleosynthesis occurred at 
$x \approx -20$.

We let the field start near the
saddle point at very early redshifts ($z \sim 10^9$) with zero
initial velocity. The resultant motion is generic and is not
sensitive to the initial values specified. In particular, at early times 
the field is 
held in place by the large Hubble friction term. Later, as the energy
density in the modulus field begins to dominate, the field starts to roll
again, providing both quintessence and a change in $\alpha$. Having
calculated the evolution of $\tau$ as a function of $z$, we can then
calculate the variation of $\alpha$ as a function of $z$, which we do
in the next section.

\section {Varying Alpha and Constraints}

The variation of $\alpha$ as a function of the bulk KK masses 
was given in Eq.~(\ref{alpha}) for one extra dimension.
This is generalized to:
\begin{equation}
\frac{\alpha_0-\alpha(z)}{\alpha_0} = \frac{\alpha_0}{2 \pi}
\tilde{b} \sum_{n_1,n_2}^{N} \frac{[{\cal M}_{0}-{\cal
      M}(z)]_{(n_1,n_2)} } { [{\cal M}_{0}]_{(n_1,n_2)} }.  
\label{final}
\end{equation}
where $\alpha_0$ and ${\cal M}_0$ are present-day values.
${\cal M}$ is implicitly a function of $z$ through its
dependence on $\tau_1(z)$.

Our ignorance of the fat brane matter content is parameterized by
the $\beta$-function coefficient; we assume $\tilde{b}$ is in the
range $1\leq\tilde b\leq 20$. For models with 2 extra dimensions with
sizes in the $\mmev^{-1}$ range, the fundamental scale of gravity $M_*$
falls
in the 10 to $100\tev$ range. The thickness of the fat brane must be
less than roughly $1\tev^{-1}$ so there is a strict upper bound on the
number of KK modes that can lie below the cutoff $M_*$. The factor $N$
which cuts off the summation in Eq.~(\ref{final}) parametrizes the
number of KK modes allowed and thus the thickness of the fat brane.
We assume $N$ is in the range $1\leq N\leq 100$. Because of modular
invariance and the need to start near the saddle point, we set
$|\tau|=1$ in our numerical work.
However we must still choose a value for $\tau_1$ at early
times; we define $\tau_0
\equiv \tau_1 (z = 10^9)$ and vary it near zero.

\begin{figure}
\centering
\hspace*{0in}
\epsfxsize=3.50in
\epsffile{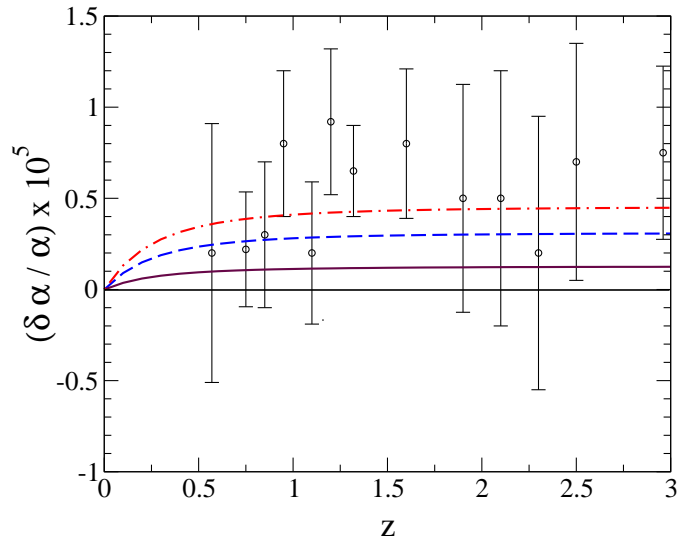}
\caption{Variation of $\alpha$ for various initial conditions and number
of KK modes ($N$) vs.\ the fiducial sample of Ref.~\cite{webb1}. The solid
line corresponds to the values $\tau_0 = 1\times10^{-5}$, 
$\tilde{b} = 15$, and $N = 2$. The dashed line corresponds to the
values $\tau_0=1\times 10^{-4}$, $\tilde{b} =4$, and $N = 8$. The
dashed-dot line corresponds to the values $\tau_0 = 1\times 10^{-3}$, 
$\tilde{b} = 1.5$, and $N = 3$.}   
\label{fig3}
\end{figure}

In Fig.~\ref{fig3} we have plotted $\frac{\delta \alpha}{\alpha}$ (in units
of $10^{-5}$) alongside the measurements of Ref.~\cite{webb1} for the
several different initial conditions and parameters given in the caption.
Note that we define $\delta \alpha(z) = \alpha_0 -\alpha(z)$. 
There are a range of initial parameters which fit the data quite
well. Each fit has a similar functional behavior, though the amplitude
can be quite different. The constraints from big bang nucleosynthesis
and WMAP are trivially satisfied because the variation in $\alpha$ is never
bigger than it is today -- that is, the change in $\alpha$ is a
relatively recent phenomenon.

Because the Oklo constraint is of a more recent origin ($z=0.14$), it
provides a highly non-trivial constraint on this model.
For example, we evaluate the variation in $\alpha$ for the 
representative examples in the figure. We find: 
\begin{equation}
\left\langle\frac{\delta\alpha}{\alpha}\right\rangle_{z=0.14} 
= (2, 6, 9) \times 10^{-7} 
\end{equation}
for the initial conditions shown in the figure. These should
be compared with with the $3\,\sigma$ bound $\delta\alpha/\alpha<
1.6 \times
10^{-7}$~\cite{dyson}. However more recent work reveals a number of
nuclear uncertainties in the calculation of the Oklo 
bound~\cite{lamoreaux}.

A separate constraint arises from measurements of the ${}^{187}$Re
decay rate in the lab and from samples taken from meteorites, which is
sensitive to redshifts back to $z\simeq 0.45$.
For our three lines, we find:  
\begin{equation}
\left\langle
\frac{\delta\alpha}{\alpha}\right\rangle_{z=0.45} = 
(5.8, 14, 21) \times 10^{-7}
\end{equation}
which should be compared with the $3\,\sigma$ bound
$\delta\alpha/\alpha <16.0\times 10^{-7}$~\cite{meteors}.

\section{Conclusions}

A cosmological time variation in the fine structure constant, if
 verified, presents a difficult  theoretical challenge. In this paper
 we have presented a new mechanism for a late-time 
 variation in a gauge coupling from time-varying charged KK modes. The
 predicted variation is ``soft," in the sense that only loop effects
 of charged Kaluza-Klein states contribute to the variation in
 $\alpha$.  This variation does not imply a change in the fundamental
 Planck scale and relies on the dimensionless shape moduli of the
 compactification manifold.  We have shown that in the case of two
 large extra dimensions it is possible to interpret both an effective
 cosmological constant and a varying $\alpha$ in terms of the dynamics
 of fields sensitive to the shape moduli. In large extra dimensional
 constructions it is natural that the Standard Model sector has a
 finite width into the bulk.  If the bulk shape moduli are dynamical
 on cosmological time scales then the Kaluza-Klein mass spectrum of
 the SM states will be altered.  However, this motion typically
 produces a large change in the vacuum energy; demanding exact
 supersymmetry above the inverse thickness scale is one possible way
 of eliminating this potentially dangerous contribution.  The only
 tuning in the model is the initial value of the moduli field at very
 early redshift.  We assume the field starts near a modular invariant
 of the torus and find that the field begins rolling near redshifts 
 $z\simeq1$.  The post-dicted variation of $\alpha$ is then 
easily reproduced for
 reasonable ranges of model parameters.

\section*{Acknowledgements}

We would like to thank I.~Gogoladze and 
P.~Regan for useful discussions. This work was supported in part by
the National Science Foundation under grant NSF--0098791, and
by a graduate fellowship
from the Notre Dame Center for Applied Mathematics (MB).

\end{document}